\documentclass[aps, prl, twocolumn, nofootinbib]{revtex4}
\usepackage{amsmath, amssymb, amsthm, graphicx}

\newcommand{\C}{\mathbb{C}}
\newcommand{\Z}{\mathbb{Z}}

\newcommand{\cK}{\mathcal{K}}
\newcommand{\cM}{\mathcal{M}}
\newcommand{\cH}{\mathcal{H}}
\newcommand{\cF}{\mathcal{F}}

\newcommand{\cO}{\mathcal{O}}
\newcommand{\cP}{\mathcal{P}}
\newcommand{\cV}{\mathcal{V}}
\newcommand{\cZ}{\mathcal{Z}}

\newcommand{\be}{\begin{equation}}
\newcommand{\ee}{\end{equation}}
\newcommand{\beq}{\begin{eqnarray}}
\newcommand{\eeq}{\end{eqnarray}}
\newcommand{\ba}{\begin{array}}
\newcommand{\ea}{\end{array}}

\newcommand{\bal}{\begin{align}}
\newcommand{\eal}{\end{align}}

\newcommand{\nsum}[1]{\sum_{#1}}

\newcommand{\pg}[3]{\phi_{g_{#1},g_{#2},g_{#3}}}
\newcommand{\bpg}[3]{\bar\phi_{g_{#1},g_{#2},g_{#3}}}

\newcommand{\pgi}[4]{\phi^{(#1)}_{g_{#2},g_{#3},g_{#4}}}
\newcommand{\pghi}[4]{\phi^{(#1)}_{g_{#2}h,g_{#3}h,g_{#4}h}}

\newcommand{\vp}{\varphi}

\newcommand{\tvp}{\tilde\varphi}

\DeclareMathOperator{\SU}{\mathrm{SU}}
\DeclareMathOperator{\tr}{\mathrm{tr}}

\newcommand{\dsty}{\displaystyle}

\newtheorem*{theorem}{Theorem}
\newtheorem*{lemma}{Lemma}
\newtheorem*{corollary}{Corollary}

\begin{document}

\title{Tensor models and embedded Riemann surfaces}
\author{James P. Ryan}\email{james.ryan@aei.mpg.de}
\affiliation{MPI f\"ur Gravitational Physics, Albert Einstein Institute, Am M\"uhlenberg 1, D-14476 Potsdam, Germany}     

\date{\today}

\begin{abstract}

Tensor models and, more generally,  group field theories are candidates for higher-dimensional quantum gravity, just as matrix models are in the $2d$ setting.   With the recent advent of a $1/N$-expansion for coloured tensor models, more focus has been given to the study of the topological aspects of their Feynman graphs.    Crucial to the aforementioned analysis  were certain subgraphs known as bubbles and jackets.   We demonstrate in the $3d$ case that these graphs are generated by matrix models embedded inside the tensor theory.  Moreover, we show that the jacket graphs represent (Heegaard) splitting surfaces for the triangulation dual to the Feynman graph.  With this in hand, we are able to re-express the Boulatov model as a quantum field theory on these Riemann surfaces.

\end{abstract}

\maketitle
Group field theories \cite{gft} and tensor models are higher dimensional analogues of matrix models \cite{mat}.  Matrix integrals have been shown to provide a natural framework within which to frame a multitude of physical and mathematical questions ranging through the fields of statistical and condensed matter physics all the way to the more abstract enumeration of virtual knots and tangles.  This highlights how apparently disparate physical and mathematical phenomena in fact share certain universal features.

One particular facet that sparked considerable interest was the realization that matrix models could give a non-perturbative definition of $2d$ quantum gravity \cite{grav}.  One considers a statistical ensemble  of $N\times N$ (oftentimes hermitian) matrices.  The Feynman graphs arising in the perturbative expansion of the free energy describe discrete Riemann surfaces. Remarkably, the expansion can be ordered in powers of $1/N$ labelled by their topological invariant, the Euler characteristic.  In the large-$N$ limit, the 2-sphere dominates and moreover,  one can tune the coupling constant so that in a double scaling limit one describes a continuum theory of 2d quantum gravity. 

Tensor models hope to reproduce the same successes that matrix models have enjoyed, with the ultimate aim of being viable candidates for quantum theories of gravity in higher dimensions. A stumbling block seemed to be that they generated a plethora of unwanted structures;  not only simplicial manifolds, but simplicial pseudo-manifolds \cite{dep}.  Recently, after much work \cite{muchw, beng}, a promising step has been made in that direction with the construction of a $1/N$-topological expansion \cite{gurN, gurriv, gurall} for the so-called {\it coloured tensor models} \cite{gurau, bos}.  In that context, it was shown that for arbitrary dimension $d$, only graphs corresponding to $d$-spheres arise at leading order in the $1/N$-expansion.  Central to this construction were the {\it ribbon graphs} associated to the Feynman graphs of the tensor model.  These ribbon graphs are algebraic objects that capture the topological properties of the Feynman graphs. They contain two classes of subgraphs, known as {\it bubbles} \cite{gurau}  and {\it jackets} \cite{beng}, which are of particular significance.  While bubbles are easily identified as Riemann surfaces embedded in the dual triangulation, the topological properties of the jackets have remained more obscure.   

We shall clarify these issues in the $3d$ scenario by identifying matrix models embedded in the tensor structure, which generate the bubbles and jackets.  

Firstly, this shows that both the bubbles and jackets may be identified with Riemann surfaces embedded in the dual triangulation. With this dual picture for the jackets we can establish some interesting properties.  In the case that the Feynman graph corresponds to a manifold, we show that the jackets correspond to Heegaard surfaces.  In the case that the Feynman graph corresponds to a pseudo-manifold, the jacket still splits the $3d$ triangulation into two handlebodies.  

Secondly, we have now recast the theories in terms of matrix models, so it opens up the avenue to analyze these models using matrix model techniques.

We shall present most of the analysis within the framework of the independent identically distributed (i.i.d.) tensor model \cite{ambj, gurall}. Near the end, we shall switch to the Boulatov model \cite{boul}, which has attracted a lot of interest since it has a manifest connection to $3d$ gravity. We shall recast this as a field theory on the jackets.  Interestingly for future work, it has a unmistakeable similarity to the  so-called dual weighted matrix models considered in \cite{kaz}.

To summarize the contents, we shall begin by describing the basic i.i.d. tensor model, after which we shall introduce the coloured formalism, including a explanation of the fundamental objects of interest, the bubbles and jackets.  We shall subsequently enter into the main part of the paper; detailing the matrix models which generate the embedded Riemann surfaces corresponding to the bubbles and jackets. We are  then in a position to establish the \lq splitting' properties of the jackets.  Before we conclude, we describe our reformulation of the Boulatov model, in the light of the previous analysis.

\section{Tensor models}
\label{tens}

\begin{figure}[h]
 \includegraphics[width = 7cm]{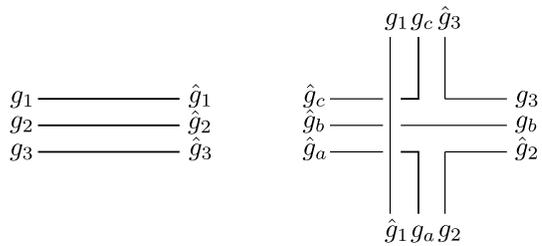}
 \caption{\label{fat} Ribbon graph components associated to elementary graph operators.}
 \end{figure}

Tensor models are higher-dimensional generalizations of matrix models.  In 3d, the fundamental object is a complex 3-tensor $\phi: \Z_N^{\times 3}\rightarrow \C$ such that $(g_1,g_2,g_3)\rightarrow \pg{1}{2}{3}$ which is subjected to some potential, for example:
\be\label{pot}
V_\lambda[\phi] =  \frac{\lambda}{4!} \nsum{g_i\in\Z_N}\mathfrak{Re}\left[ \pg{c}{b}{a}\,\pg{1}{a}{2}\,\pg{2}{b}{3}\,\pg{3}{c}{1}\right]
\ee
 We consider the free energy of a statistical ensemble of such tensors with respect to the potential \eqref{pot}:
\be
\cF_\lambda = \ln \int d\mu[\phi]\; e^{V_\lambda[\phi]} =  \sum_{\Gamma}\frac{1}{|Aut[\Gamma]|}\cZ_{\lambda,\Gamma}.
\ee
We have used a normalized Gaussian measure on the space of complex 3-tensors:
\be
d\mu[\phi]  = \frac{\dsty\prod_{g_{i}} \Big[d\mathfrak{Re}[\phi_{g_i}]\;d\mathfrak{Im}[\phi_{g_i}]\Big]  e^{-\sum_{g_i}|\phi_{g_i}|^2 }  }
 {\dsty\int \prod_{g_{i}} \Big[d\mathfrak{Re}[\phi_{g_i}]\;d\mathfrak{Im}[\phi_{g_i}]\Big]  e^{-\sum_{g_i}|\phi_{g_i}|^2}  }
\ee
with the Lebesgue measure for each of the tensor components.  In the perturbative expansion,  $\Gamma$ are the connected 4-valent Feynman graphs, $|Aut[\Gamma]|$ is the order of the discrete automorphism group of $\Gamma$ and $\cZ_\Gamma$ is the amplitude associated to $\Gamma$, which we shall presently construct. From a field theoretic perspective,  one associates to the graph the following propagator (coming from the measure) and vertex operator:
\be
\ba{rcl}
\cP_{g_i;\bar g_i} &=&\delta_{g_1,\hat g_1}\,\delta_{g_2,\hat g_2}\,\delta_{g_3,\hat g_3}	\\[0.2cm]
\cV_{g_i;\hat g_i} &=&\dfrac{\lambda}{4!}\delta_{g_1, \hat g_1 }\,\delta_{g_2, \hat g_2 }\,\delta_{g_3, \hat g_3 }\delta_{g_4, \hat g_4 }\,\delta_{g_5, \hat g_5 }\,\delta_{g_6, \hat g_6}
\ea
\ee
 where $\delta$ is the Kronecker-$\delta$ on $\Z_N$.  Supplementing this with the obvious summation of group variables, one has the complete set of Feynman rules.  But the peculiar coupling of tensor components in the potential hints that the theory knows more about the topological structure of the Feynman diagram than just its 1-skeleton. One would like to make this property more transparent. To that effect,  one \lq fattens' each Feyman diagram $\Gamma$ to an associated {\it ribbon graph} $r[\Gamma]$.  More precisely, one replaces each of its lines with three strands which are re-routed at the vertex according to the Kronecker-$\delta$ weights, see Fig \ref{fat}.
 
\noindent With $r(\Gamma)$ at our disposal, we see that the tensor theory knows about the 2-skeleton of $\Gamma$, that is,  the vertices, edges and faces, which are just the closed loops in $r(\Gamma)$.
\be
\cZ_{\lambda, \Gamma} = \lambda^{|v_\Gamma|}N^{|f_\Gamma|}  = \cZ_{\Gamma, DT}
\ee
 with $|v_\Gamma|$ and $|f_\Gamma|$ representing the total number of vertices and faces in $\Gamma$, respectively.   This amplitude may be re-interpreted as coming from the quantization of gravity in the dynamical triangulation (DT) approach.  One interprets the Feynman graph as the dual to an equilateral 3d triangulation $\Delta$. We denote the vertices, edges, triangles and tetrahedra of $\Delta$ as $v_\Delta$, $e_\Delta$, $f_\Delta$ and  $t_\Delta$, while the vertices, edges, faces and 3-cells of $\Gamma$ are denoted by $v_\Gamma$, $e_\Gamma$, $f_\Gamma$ and $b_\Gamma$.  The two structures are dual in the sense that there is a one-to-one correspondence: $v_\Gamma\sim t_\Delta$,  $e_\Gamma\sim f_\Delta$, $f_\Gamma\sim e_\Delta$ and $b_\Gamma\sim v_\Delta$.  Then, by substituting $N = \exp(1/G)$ and $\lambda = \exp(-(1+c)/G)$ the amplitude takes the form:
 \be
 \cZ_{\Gamma, DT} = e^{-\Lambda |v| + \frac{1}{G}|f| - \frac{c}{G}|v|  }
 \ee
where $\Lambda$ and $G$ are the bare cosmological and gravitational constants respectively. One takes all the tetrahedra to be of unit volume and therefore $|v_\Gamma|$, the number of tetrahedra,  is the volume of the $\Delta$.   Finally, $c$ is a geometric constant arising from the particular Regge discretization of the action inherent in the DT approach. 
A similar model to that above was proposed and studied in \cite{ambj} (the 3-tensors satisfied a condition analogous to hermiticity in matrix models,  $\bpg{1}{2}{3} = \pg{3}{2}{1}$, along with invariance under even permutations).   The authors ultimately drew a number of negative conclusions.  Notably, the potential is fourth order and  unbounded from below. Therefore, the free energy is divergent for all $\lambda\neq 0$ and the perturbative expansion is at most a formal object.  Moreover,  the model does not have a well-behaved large-$N$ limit: $1/G$ diverges as $N\rightarrow \infty$.  They also expressed concern that the Feynman graphs are not generically dual to simplicial manifolds but to a more general class of objects known as simplicial pesudo-manifolds.  Recently, more light has been shed on these issues with the advent of {\it coloured group field theories}.

%


\section{Coloured tensor models}
\label{col}

A nifty addition to the tensor programme, known as {\it colouring}, has recently been proposed \cite{gurau}.  One replaces the complex 3-tensor of the previous section with four such objects $\phi^{(i)}: \Z_N^{\times 3}\rightarrow \C$, $i\in\{0,1,2,3\}$, subjected as usual to one's favorite (coloured) potential, for example:
\be\label{col01}
\ba{l}
\dsty
V_\lambda[\phi^{(i)}] =  \frac\lambda {\sqrt{N^3}} \nsum{g_i\in\Z_N}\mathfrak{Re}\left[ \pgi{0}{c}{b}{a}\,\pgi{1}{1}{a}{2}\right.\\[0.2cm]
\hspace{4.5cm}\left.\times\;\pgi{2}{2}{b}{3}\,\pgi{3}{3}{c}{1}\right]
\ea
\ee
With respect to a Gaussian measure on each of the fields, one finds that the Feynman amplitudes of this model are:
\be\label{col01a}
\cZ_{\Gamma} = \lambda^{|v_\Gamma|}N^{|f_\Gamma| - \frac12|v_\Gamma|}  
\ee
where $|v_\Gamma|$ and $|f_\Gamma|$ are the total number of vertices and faces, summing over colours. They only change is the rescaling of the coupling constant. The boon is that one has more control over the type of diagram arising.  

\subsubsection{Boundedness of the potential:}
To commence, let us investigate the boundedness of the potential.  At first sight, the situation might seem even more hopeless, since we have four independently fluctuating fields.  But, at least for the free energy, one can integrate out the $\phi^{(0)}$-field to get an effective potential:
\be\label{col02}
V_\lambda[\phi^{(i)}] =  \frac{\lambda^2}{N^3} \sum_{g_{a,b,c}}\left| \sum_{g_{1,2,3}}\pgi{1}{1}{a}{2}\,\pgi{2}{2}{b}{3}\,\pgi{3}{3}{c}{1}\right|^2
\ee 
for the remaining fields that is manifestly bounded from below.   Thus, one may hope that the perturbative expansion is better defined than for the non-coloured case.

\subsubsection{Scaling of the coupling constant:}

\begin{figure}[h]
\includegraphics[width=4cm]{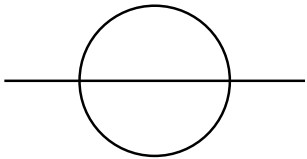}
\caption{\label{topo} Topologically trivial edge insertion.}
\end{figure}

The importance of the fact that the coupling constant scales as $\lambda/\sqrt{N^3}$ has been stressed in \cite{beng} and subsequently in \cite{gurall}, where the relevant details are expressed.  To explain briefly, consider two graphs $\Gamma_1$ and $\Gamma_2$  in the expansion, such that $\Gamma_2$ is just $\Gamma_1$  supplemented with a single insertion of Fig. \ref{topo} along some edge.  $\Gamma_1$ and $\Gamma_2$ have the same topology but for a general scaling $\lambda/N^\alpha$, the amplitudes are related by $\cZ_{\Gamma_2} = (\lambda^2/N^{2\alpha - 3}) \cZ_{\Gamma_1}$.  Thus, it is only for $\alpha = 3/2$ that the $1/N$-topological expansion could make sense.

\subsubsection{Tracking information}

Since we are dealing with coloured complex fields and the potential given above, we find that the vertices, edges, faces and 3-cells of $\Gamma$ come in a variety of types. The vertices come in two types, corresponding to the two potential terms.    The edges are maximally connected subgraphs of one colour and thus come in four types $e_\Gamma^{(i)}$, where $i\in\{0,1,2,3\}$.  This is evident because the fields are coloured.   Faces are maximally connected subgraphs of two colours. They are formed from loops of edges alternating between any two colours, that is, they come in six types $f_\Gamma^{(ij)}$, where $i,j\in\{0,1,2,3\}$ and $i\neq j$.    The 3-cells are maximally connected subgraphs of three colours.  They come in four varieties depending on which colour is {\it not} present.  Thus, the 3-cells $b_\Gamma^{(i)}$ are formed by deleting all edges $e_\Gamma^{(i)}$ from $\Gamma$.   

As one might imagine, the sub-cells of the dual triangulation inherit the colouring. Thus, each  3-cell $b_\Gamma^{(i)}$ encloses a vertex $v_\Delta^{(i)}$, each face $f_\Gamma^{(ij)}$ loops around an edge  $e_\Delta^{(ij)}$ and each edge  $e_\Gamma^{(i)}$ pierces a triangle $f_\Delta^{(i)}$.

\subsubsection{Bubbles and Jackets}

The {\it bubbles} $B_\Gamma^{(i)}$ of $\Gamma$  \cite{gurau} are defined as the maximally connected subgraphs of $r[\Gamma]$ with three colours and are thus the ribbon graphs associated to the 3-cells $b_\Gamma^{(i)}$, that is, $B_\Gamma^{(i)} = r[b_\Gamma^{(i)}]$.  

A {\it jacket} $r[J^{(ij,\hat{ij})}]$ of $\Gamma$ \cite{beng} is defined as the ribbon graph obtained from $r[\Gamma]$ by deleting all the faces $f_\Gamma^{(ij)}$ and $f_\Gamma^{(\hat{ij})}$ where $\hat{ij} = \{0,1,2,3\}\backslash \{i,j\}$.  There are clearly three jackets for each $\Gamma$ depending on whether one chooses $(ij) = (01)$, $(02)$ or $(03)$. (The other three choices are equivalent to one of those mentioned).

\begin{figure}[h]
\includegraphics[width = 8cm]{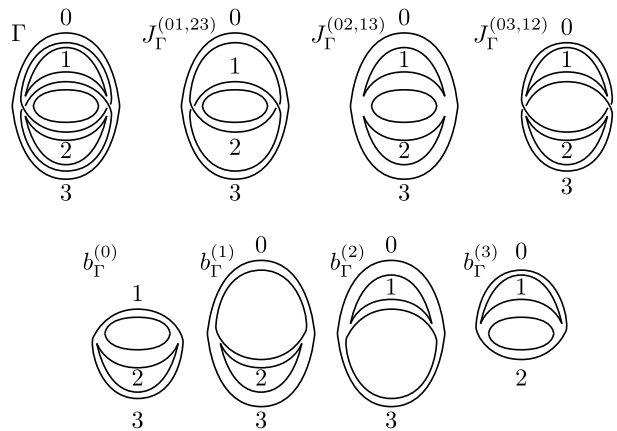}
\caption{\label{bubintro} $\Gamma$ with its bubbles and jackets}
\end{figure}

We illustrate using the traditional and simplest example in Fig. \ref{bubintro}.

We wish to highlight that all the ribbon graphs appearing here have two strands.  One knows well that these are exactly the kind of graphs arising in the study of matrix models.  Later, we shall identify the matrix models embedded in the tensor model that generate exactly these bubbles and jackets.

\subsubsection{$1/N$-expansion and large-$N$ behaviour}

Recently, Gurau identified certain {\it core graphs}  which are homeomorphic to the original $\Gamma$ and entirely encode their scaling with respect to $N$ \cite{gurN}.      We shall not present the details here, but the results of his analysis establish that at leading order in $1/N$ only graphs corresponding to the $3$-sphere arose.    This result was extended to arbitrary dimension \cite{gurriv} and later more refined details on the suppressed terms were uncovered \cite{gurall}.

While a $1/N$-expansion is now possible, the large $N$ behaviour of this particular model is still not suited to describing gravity for the same reason as in the non-coloured case.  Luckily this analysis has been also completed for the Boulatov model, which has a less opaque connection to $3d$ quantum gravity. Here, we shall focus on investigating the bubbles and jackets.


\section{Revisiting bubbles and jackets}

The algebraic objects of the previous section have a fundamental importance in classifying the properties of tensor models.  The topological properties of the bubbles are fairly transparent; they correspond to embedded Riemann surfaces surrounding vertices of $\Delta$.  We shall see that for each colour, there is an embedded matrix model, which generates bubbles of that colour.  Furthermore,   we shall perform the same analysis for the jackets, attempting in the process to make their topological properties as manifest as possible, with the hope that it will provide yet another tool with which to tackle this class of theories.

\subsection{Bubbles}

For this analysis, we shall first perform the perturbative expansion, integrating with respect to $\phi^{(1,2,3)}$ and leaving $\phi^{(0)}$ untouched.  To illuminate our reasoning, let us reinterpret the tensors $\phi^{(1,2,3)}$ as a set of $3N$  complex $N\times N$ matrices: $[\Phi^{(1,2,3)}_{g_a}]_{g_1,g_2} = \phi^{(1,2,3)}_{g_1,g_a,g_2}$. Moreover, we shall briefly view $t = \phi^{(0)}$ as a coupling parameter. Thus, the potential \eqref{col01} takes the form:
\be\label{col03}
V_\lambda[t,\Phi^{(i)}] =  \frac{\lambda}{\sqrt{N^3}} \sum_{g_{a,b,c}} \mathfrak{Re}\left[ t_{g_{c},g_{b},g_{a}} \tr(\Phi^{(1)}_{g_a}\,\Phi^{(2)}_{g_b}\,\Phi^{(3)}_{g_c})\right],
\ee
where the trace is over the $N\times N$ matrices. The free energy can be re-expressed as:
\be\label{col04}
\cF_\lambda = \int d\mu[t]\,  e^{\cF_\lambda[t]}
\ee
where $\cF_\lambda[t]$ is the free energy of the matrix model with potential \eqref{col03}:
\be\label{col05}
\cF_\lambda[t]= \ln \int d\mu[\Phi^{(i)}]\; e^{V_\lambda[t,\Phi^{(i)}]} = \sum_{b^{(0)}} \frac{\cZ_{b^{(0)}, \lambda}[t]}{|Aut[b^{(0)}]|},
\ee
The amplitude associated to a Feynman graph ${b^{(0)}}$ of this matrix model is:
\be\label{col06}\ba{rcl}
\cZ_{{b^{(0)}},\lambda}[t]&=& \dsty \lambda^{|v_{b^{(0)}}|}\, N^{|f_{b^{(0)}}|  -\frac{3}{2}|v_{b^{(0)}}|}\, \cO_{b^{(0)}}[t]
\ea
\ee
where $|v_{b^{(0)}}|$ and $|f_{b^{(0)}}|$ are, respectively, the total number of vertices and faces of ${b^{(0)}}$.  By construction,  we have a multi-matrix model with cubic potentials. Thus, each Feynman graph is a trivalent graph dual to a triangulation of a connected orientable Riemann surface.  As expected, one gets a factor of $N$ for every face of ${b^{(0)}}$ and $\lambda/\sqrt{N^3}$ for ever vertex of ${b^{(0)}}$. The final factor $\cO_{b^{(0)}}[t]$ is a polynomial in $t$ and $\bar{t}$ based on the graph ${b^{(0)}}$. In effect, $t$ and $\bar{t}$ label the vertices of ${b^{(0)}}$, while their components label the edges: 
\be
\cO_{b^{(0)}}[t] = \left[\prod_{e_{b^{(0)}}}\sum_{g_{e_{b^{(0)}}}}\right]\left[ \prod_{v_{b^{(0)}}} t_{g_{e_{b^{(0)}}@ v_{b^{(0)}}}}\right],
\ee
where $e_{b^{(0)}}$ are the edges of ${b^{(0)}}$ and $e_{b^{(0)}}@v_{b^{(0)}}$ are the three edges incident at $v_{b^{(0)}}$. There is a further admissibility condition: for a vertex labelled by $t$, the adjacent vertices must be labelled by $\bar{t}$ and vice versa. This observable in $t$ is very familiar from the spin foam framework; it is the evaluation of a product of tensors assigned to the (spin-network) graph ${b^{(0)}}$.

At this intermediate position, let us say a word or two on the role these Riemann surfaces play within the context of the tensor model.  Note that in moving to the matrix model \eqref{col03}, we have re-interpreted the strands coupling $\phi^{(0)}$ to $\phi^{(1)}$, $\phi^{(2)}$ and $\phi^{(3)}$ as different species of matrix.  Therefore, they are deleted from the ribbon graph of the tensor model to get the ribbon graph of the matrix model.  But these are exactly, the bubbles $B^{(0)} = r[b^{(0)}]$.

\begin{figure}[h]
\includegraphics[width = 8cm]{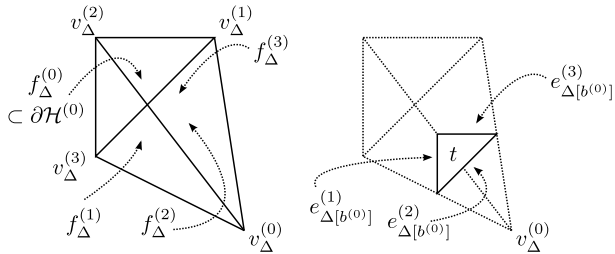}
\caption{\label{bub}The correspondence between the tensor and matrix pictures.}
\end{figure}

Let us describe $\cZ_{b^{(0)},\lambda}[t]$ from the dual $2d$ and  $3d$ perspectives. Remember that $\Gamma$ is dual to a triangulation $\Delta$ and we shall denote the dual to the $b^{(0)}$ as $\Delta[ b^{(0)}]$.    In the $2d$ setting, the matrix model glues collections of triangles to form triangulated Riemann surfaces $\Delta[ b^{(0)}]$ whose triangles are weighted by the tensor $t$ and have edges coloured $e_{\Delta[b^{(0)}]}^{(1)}$, $e_{\Delta [b^{(0)}]}^{(2)}$ and $e_{\Delta[ b^{(0)}]}^{(3)}$. Remember that integrating with respect to $\Phi^{(1,2,3)}$ in the $2d$ picture is equivalent to integrating with respect to $\phi^{(1,2,3)}$ in the $3d$ picture.  In the $3d$ setting, one has a collection of tetrahedra, with triangles coloured $f_\Delta^{(0)}$, $f_{\Delta}^{(1)}$, $f_\Delta^{(2)}$ and $f_\Delta^{(3)}$.    As a result, one glues tetrahedra along triangles of colour $(1)$, $(2)$ and $(3)$, leaving the triangles of colour $(0)$ open. This generates a handlebody $\cH^{(0)}$, the boundary of which is a triangulated Riemann surface $\partial \cH^{(0)}$ identical to $\Delta[ b^{(0)}]$.   The vertices of $\partial \cH^{(0)}$ are all of type $v_\Delta^{(1)}$, $v_\Delta^{(2)}$ and $v_\Delta^{(3)}$, while there is one interior vertex enclosed by $\partial\cH^{(0)}$ (and $\Delta[b^{(0)}]$) of type $v_{\Delta}^{(0)}$.  See Fig. \ref{bub} for an illustration.  

%

To calculate the free energy of the tensor theory \eqref{col04}, we must integrate the {\it partition function} of the matrix model with respect to the couplings $t$.  In the matrix model partition function, we shall have multi-component graphs or in other words, collections of bubbles $\{B^{(0)}\}$.  Integrating with respect to $t$ reintroduces strands among the bubbles necessary to reconstruct a $3d$ ribbon  graph $r[\Gamma]$.   

Once again, from the $3d$ dual perspective, after the first step, we generated triangulated handlebodies $\cH^{(0)}$.  A typical contribution to the tensor model free energy comes from gluing a collection of handlebodies $\{\cH^{(0)}\}$. Integrating with respect to the measure $\mu[t]$ essentially completely glues these handlebodies together to form a closed 3d triangulation.  The gluing procedure is completely standard from the topological point of view.  One repeatedly identifies  pairs of discs (in this case triangles) on the boundaries of the handlebodies and forms the connected product.   Of course, in the end, one can index these $3d$ structures by their constituent handlebodies and the gluing maps.

To describe the reconstruction of the amplitude for $\Gamma$ in words is somewhat cumbersome, but it is possible to keep track of the factors of $N$ and arrive back at \eqref{col01a}.

Note that there are four distinct redefinitions of the form \eqref{col03},  yielding matrix models that generate $b^{(1)}$, $b^{(2)}$ and $b^{(3)}$ graphs, respectively, at the intermediate stage. 

Although pseudo-manifolds are suppressed in the $1/N$-expansion of this model, one might have retained some hope that they could be removed completely by some restriction on the tensor model.  Our analysis here serves to highlight just how drastic a restriction this would be.  In order for a triangulation to be a manifold, one must ensure that all bubbles (of every colour) are spherical.  But one sees that bubbles arise from matrix models embedded inside the tensor model.  Thus higher order bubble topologies are abundant and completely natural from this point of view.  In fact, to restrict sharply to just the spherical topology is in many ways the antithesis of matrix model ideology.

\subsection{Jackets}

\begin{figure}[h]
\includegraphics[width = 8cm]{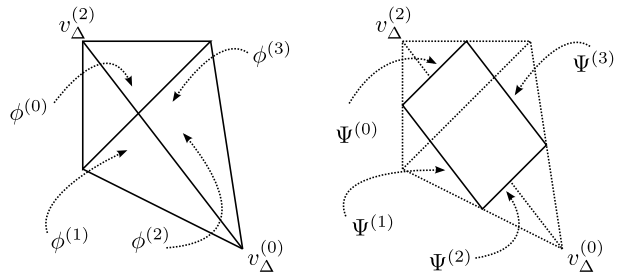}
\caption{\label{jacp} The correspondence between the tensor and matrix descriptions.}
\end{figure}

To make the construction of jackets manifest at the level of the action, one chooses a different redefinition.  We reinterpret the tensors $\phi^{(0,1,2,3)}$ as a set of $4N$ complex $N\times N$ matrices in the following way:
\be\label{jac01}
\ba{rclcrcl}
[\Psi^{(0)}_{g_a}]_{g_1, g_2} &=& \phi^{(0)}_{g_1,g_a,g_2},&  [\Psi^{(1)}_{g_a}]_{g_1, g_2} &=& \phi^{(1)}_{g_a,g_1,g_2},\\[0.3cm]
[\Psi^{(2)}_{g_a}]_{g_1, g_2} &=& \phi^{(2)}_{g_1,g_a,g_2},&  [\Psi^{(3)}_{g_a}]_{g_1, g_2} &=& \phi^{(3)}_{g_1,g_2,g_a}.
\ea
\ee
With this redefinition, the potential \eqref{col01} takes the form:
\be
V_\lambda[\Psi^{(i)}] = \frac\lambda{\sqrt{N^3}} \sum_{g_a,g_b} \mathfrak{Re} \left[\tr(\Psi^{(0)}_{g_a}\,\Psi^{(1)}_{g_b}\,\Psi^{(2)}_{g_a}\,\Psi^{(3)}_{g_b})\right]
\ee
Clearly, we have a complex multi-matrix model, albeit a very unusual one, since the number of species of matrix is coupled to the size of the matrices. In the perturbative expansion of the free energy, one generates 4-valent graphs $J$ dual to quadrangulations of connected oriented Riemann surfaces. Extra degrees of freedom propagate along the surface, corresponding to the multitude of species of matrix.  In any case, note that it is the strand coupling $\phi^{(0)}$ and $\phi^{(2)}$ and the strand coupling $\phi^{(1)}$ to $\phi^{(3)}$ that become the various species of matrices $\Psi^{(i)}_g$.  They are deleted from the ribbon graph $r[\Gamma]$ of the tensor model to get the ribbon graphs $r[J]$ of the matrix model above.  As we anticipated, this matrix model generates exactly the jackets $r[J^{(02,13)}]$ and we see that it is dual to a Riemann surface embedded inside $\Delta$, see Fig. \ref{jacp}.

The amplitude for a given jacket takes the form:
\be
\cZ_{\lambda,J^{(02,13)}} = \lambda^{|v_J|} N^{|f_J| - \frac12|v_J| + |c_J|},
\ee
where  $|v_J|$ and $|f_J|$ are the total number of vertices and faces of the Feynman graph. Note that the vertices of $J$ are in one-to-one correspondence with the vertices of $\Gamma$, while the faces of $J$ are in one-to-one correspondent with the faces of $\Gamma$ which are not of colour $(02)$ or $(13)$.  From this $2d$ perspective, $|c_J|$ arises because we have $4N$ species of matrix.  But of course we already know that it is  the cardinality of the set of loops we deleted from the ribbon graph $r[\Gamma]$ to get $r[J^{(02,13)}]$:  $|c_J| = |f_\Gamma| - |f_J|$.  

To finish, just as there are three jackets for each graph $\Gamma$, there are three distinct redefinitions of the form \eqref{jac01} at the level of the action, each one generating a different jacket.

\subsection{Jackets as Heegaard surfaces}

Here we shall show that the Riemann surfaces corresponding to the jackets are in fact Heegaard surfaces.  

Formally, a {\it Heegaard splitting} of a compact connected oriented 3-manifold $\cM$ is an ordered triple $(\Sigma, \cH_1, \cH_2)_\cM$. $\Sigma$ is a compact connected oriented surface, while $\cH_1$ and $\cH_2$ are handlebodies.  All three are embedded in $\cM$ such that  $\Sigma = \partial \cH_1 = \partial \cH_2$.  $\Sigma$ is known as a {\it Heegaard surface}.

A {\it spine} of a handlebody $\cH$ is a (piecewise linear) graph $\cK$ embedded in $\cH$ such that $\cH\backslash\cK$  is homeomorphic to $\partial \cH\times (0,1]$. Moreover, if one has a piecewise linear graph $\cK$ and $\cH$ is the closure of a regular neighborhood of $\cK$, then $\cK$ is a spine of $\cH$.

Let us consider the coloured triangulation dual to $\Gamma$ and define   $\cK^{(ij)} = \{v_\Delta^{(i)}, v_\Delta^{(j)}, e_\Delta^{(ij)}\}$, that is, the set of all  $(i)$-vertices, all $(j)$-vertices and all $(ij)$-edges in $\Delta$.

{\lemma  $\cK^{(ij)}$ is a connected piecewise linear graph in $\Delta$.  $\cK^{(ij)}$ and $\cK^{(\hat{ij})}$ are disjoint graphs.    }

This is rather evident from the construction of the 3d triangulation.  

 Let us define $\square[J^{(ij, \hat{ij})}]$ as the quadrangulation corresponding to the jacket $r[J^{(ij,\hat{ij})}]$ and $\cH^{(ij)}$ as the closure of a neighbourhood of $\cK^{(ij)}$.

{\theorem If $\Delta$ is a manifold, then the triple $(\square[J^{(ij, \hat{ij})}], \cH^{(ij)}, \cH^{\hat{(ij)}})_\Delta$ is a Heegaard splitting  of $\Delta$.}

{\proof $\cH^{(ij)}$ is defined as the closure of a neighbourhood of $\cK^{(ij)}$ in $\Delta$. Let us examine, the part of this  neighbourhood inside a given tetrahedron. 

\begin{figure}[h]
\includegraphics[width = 8cm]{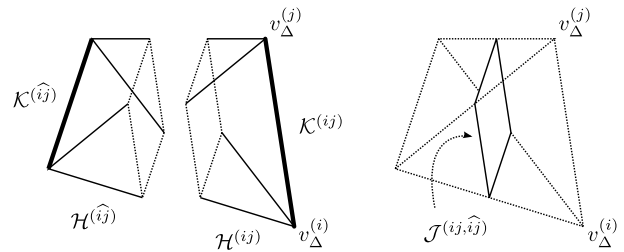} 
\caption{\label{heeg} The interection of the spines, handlebodies and splitting surface with a single tetrahedron.}
\end{figure}

As illustrated in Fig. \ref{heeg}, the neighbourhoods of $\cK^{(ij)}$ and $\cK^{(\widehat{ij})}$  can be extended as far as some intermediary surface, which is homeomorphic to the quadrangular disc in $\square[J^{(ij,\hat{ij})}]$.  Furthermore, gluing tetrahedra around an edge of $\Delta$ causes no problems.  As expected, however, problems could arise at the vertices of $\Delta$.   To see this, let us examine Fig. \ref{heegvert}.
\begin{figure}[h]
\includegraphics[width = 3cm]{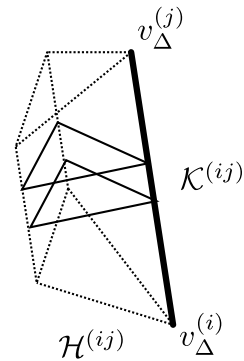}
\caption{\label{heegvert} Splitting $H^{(ij)}$ into three parts in each tetrahedron.}
\end{figure}
Within each tetrahedron, we split $\cH^{(ij)}$ into three segments, one which touches $v^{(i)}_\Delta$, one which touches $v_\Delta^{(j)}$ and one which touches neither, just the edge $e_\Delta^{(ij)}$.  Upon gluing the tetrahedra, the segments which touch a given $v^{(i)}_\Delta$ glue to become a handlebody in $\Delta$ with boundary homeomorphic to the bubble handlebody $\cH^{(i)}$ enclosing $v_\Delta^{(i)}$.  Meanwhile,  after gluing the tetrahedra, the segments touching just $e_\Delta^{(ij)}$ form a fat disc.

With this decomposition of $\cH^{(ij)}$,  it is easier to see that if one attempts to perform a deformation retraction of $\cH^{(ij)}$ onto $\cK^{(ij)}$, one hits an obstruction unless all the handlebodies enclosing the vertices are balls, in which case $\Delta$ is a manifold. When $\Delta$ is a manifold,  the deformation retraction can be performed, $\cH^{(ij)}$ is a regular neighbourhood of $\cK^{(ij)}$ and  $\cH^{(ij)}$ is a handlebody with $K^{(ij)}$ as a spine.  The same holds for $\cH^{(\widehat{ij})}$.  Moreover,  $\square [J^{(ij,\hat{ij})}] = \partial \cH^{(ij)} = \partial \cH^{(\hat{ij})}$, from which the result follows. 

When $\Delta$ is a pseudo-manifold, we do not find that $\cK^{(ij)}$ is a spine for $\cH^{(ij)}$.  It is clear from the argument,  however, that $\cH^{(ij)}$ is a collection of handlebodies connected by solid tubes satisfying $\square [J^{(ij,\hat{ij})}] = \partial \cH^{(ij)} = \partial \cH^{(\hat{ij})}$.  Thus $\square [J^{(ij,\hat{ij})}] $ splits $\Delta$ into  two handlebodies $\cH^{(ij)}$ and $\cH^{(\hat{ij})}$.

}

It is a rather straightforward to show a relation between the Euler character of the jackets and the bubbles.  

{\corollary The following relation holds:
$$\chi(\square [J^{(ij,\hat{ij})}] ) = \sum_{b^{(i)}} \chi(\Delta[b^{(i)}]) +  \sum_{b^{(j)}} \chi(\Delta[b^{(j)}]) - 2|e_\Delta^{ij}|$$.}

\section{ $\Z_N$ Boulatov model}

Interest in tensor models was first re-ignited with Boulatov's modification \cite{boul}.  Significantly, the space of fields occurring in \eqref{col01} is projected down to those invariant under the following symmetry:
\be\label{sym}
\pgi{i}{1}{2}{3} = \pghi{i}{1}{2}{3} \quad\textrm{for all}\quad  h\in\Z_N,
\ee
known as a diagonal shift-symmetry.  We can impose this symmetry by explicitly averaging over non-invariant fields $\tilde \phi$:
\be
\phi^{(i)}_{g_1,g_2,g_3} = \frac{1}{N}\sum_h \tilde \phi^{(i)}_{g_1h,g_2h,g_3h}.
\ee
 This modifies the operators to:
\be
\ba{rcl}
\cP_{g_i;\hat g_i} &=&\dsty \frac{1}{N}\sum_{h}\delta_{g_1h,\hat g_1}\,\delta_{g_2h,\hat g_2}\,\delta_{g_3h,\hat g_3}	\\[0.4cm]
\cV_{g_i;\hat g_i} &=&\dsty \frac{\lambda}{\sqrt{N}N^4}\sum_{h_{a,b,c,d}}
\delta_{g_1h_a, \hat g_1 h_d}\,\delta_{g_2h_a, \hat g_2 h_c}\,\delta_{g_3h_a, \hat g_3 h_b}\\[0.2cm]
&& \phantom{xxxxxxxx}\times\delta_{g_4h_b, \hat g_4 h_c}\,\delta_{g_5h_b, \hat g_5 h_d}\,\delta_{g_6h_c, \hat g_6 h_d} 
\ea
\ee
One notes that there are a number of $\Z_N$-averages taking place in the above operators and this reflects the property \eqref{sym}. Moreover, the appropriate scaling of the coupling constant in this case is $\lambda/\sqrt{N}$ \cite{beng}.    The graph amplitude now takes the form:
\be
\cZ_{\Gamma} = \frac{\lambda^{|v_{\Gamma}|}}{N^{|e_\Gamma| + \frac{1}{2}|v_\Gamma|}} \left[\prod_{e^{(i)}_\Gamma} \sum_{h_{e}}\right] \prod_{f^{(ij)}_\Gamma}\delta\left( \sum_{e\subset\partial f} h_{e}^{\epsilon(e,f)} \right)
\ee
where $|v_\Gamma|$ and  $|e_\Gamma|$  are the total number of vertices and edges of $\Gamma$ of {\it all} colours and $\epsilon(e,f) = \pm 1$ depending on the relative orientation of $e^{(i)}_\Gamma$ and $f^{(ij)}_{\Gamma}$.  Thinking of the group elements $h_e$ as representing a $\Z_N$-valued connection on $\Gamma$, then the face weight just enforces the $\Z_N$-flatness of this $\Z_N$-connection.  Such an amplitude arises in the quantization of $\Z_N$ $BF$ theory on a triangulation \cite{bdr}.  

One can simplify the amplitude \cite{boul}:
\be\label{div}
\cZ_{\Gamma, BF} = N^{|b_\Gamma|  - \frac{1}{2}|v_\Gamma| - 1 + b_2[\Gamma, \Z_N]}
\ee
where $b_2[\Gamma, \Z_N] = \mathrm{rank}(H_2[\Gamma, \Z_N])$ is the rank of the second homology group on $\Gamma$ with coefficients in $\Z_N$ \cite{hatch}.  Thus, the $BF$ amplitude is almost a topological invariant.   Its relation to gravity comes about when one changes to a field $\phi: \SU(2)^{\times 3}\rightarrow \C$.  The resulting $\SU(2)$ $BF$ amplitude is known to arise as a quantization of a first order form of 3d gravity on the graph $\Gamma$.  Furthermore, moving to a non-abelian (Lie) group, introduces many subtleties in placing the amplitude in a form similar to \eqref{div}, but this has been successfully accomplished in a series of works \cite{bfdiv}.

\subsection{Jacket field theory}

We would like to express the theory as a matrix model for the jacket, but at the outset the symmetry \eqref{sym} seems to spoil this possibility.  To circumvent this problem, we start by the gauge-fixing the symmetry.  It turns out that there are several gauge-fixings which are trivial in the sense that one can show easily that one obtains the same graph amplitudes in the perturbative expansion. These gauge-fixings have been used before \cite{livgir}, but we shall take them in a new direction. First, the symmetry essentially projects the domain of the field $\phi$ onto $\Z_N^{\times 2}$, which we shall make explicit by introducing  new fields $\varphi^{(i)}:\Z_N^{\times 2}\rightarrow \C$ such that:
\be\label{jft01}
\ba{rclcrcl}
\varphi^{(0)}_{g_1g_a^{-1}, g_2g_{a}^{-1}} &=& \phi^{(0)}_{g_1, 1,g_2},&  \varphi^{(1)}_{g_1g_a^{-1}, g_2g_{a}^{-1}} &=& \phi^{(1)}_{1,g_1,g_2},\\[0.3cm]
\varphi^{(2)}_{g_1g_a^{-1}, g_2g_{a}^{-1}} &=& \phi^{(2)}_{g_1,1,g_2},&  \varphi^{(3)}_{g_1g_a^{-1}, g_2g_{a}^{-1}} &=& \phi^{(3)}_{g_1,g_2,1}.
\ea
\ee
Note the similarity between this redefinition and that in \eqref{jac01}.   In essence, we use the symmetry to fix one of the tensor components of each of the fields, in such a way that only degrees of freedom propagate along the jackets.   The potential now takes the form:
\be\label{boupot}
\ba{l}
\dsty
V_\lambda[\varphi^{(i)}] =  \frac\lambda {\sqrt{N^3}} \nsum{g, g_i}\mathfrak{Re}\left[ \varphi^{(0)}_{g_1,g_2}\,\varphi^{(1)}_{g_2g,g_3g}\right.\\[0.2cm]
\hspace{4.5cm}\left.\times\;\varphi^{(2)}_{g_3,g_4}\,\varphi^{(3)}_{g_3g,g_1g}\right]
\ea
\ee 
and the Gaussian measure on each of the fields is:
\be
d\mu[\varphi^{(i)}]  = \frac{1}{c[\varphi^{i}]}\prod_{g_{a}} \Big[d\mathfrak{Re}[\varphi^{(i)}_{g_a}]\;d\mathfrak{Im}[\varphi^{(i)}_{g_a}]\Big]  e^{-\sum_{g_a}|\varphi^{(i)}_{g_a}|^2 }
\ee
where $c[\varphi^{(i)}]$ is the appropriate normalization factor.\footnote{Note that we rescaled the fields $\varphi^{(i)}\rightarrow \varphi^{(i)}/\sqrt{N}$ to put the potential and measure in that form.  This leads to the $\lambda/\sqrt{N^3}$ factor.} Although we have it in a purely matrix format, we would like to represent potential as a trace.  This can be accomplished if we perform a discrete Fourier decomposition on the tensor components:
\be
\ba{l}
\dsty\vp^{(0,2)}_{g_1,g_2} =  \sum_{x_i\in\Z_N} \tvp^{(0,2)}_{-x_1,-x_2}\; e^{\frac{2\pi i}{N} (x_1g_1  + x_2g_2)}\\[0.4cm]
\dsty\vp^{(1,3)}_{g_1,g_2} = \sum_{x_i\in\Z_N} \tvp^{(1,3)}_{x_1,x_2}\; e^{\frac{2\pi i}{N} (x_1g_1  + x_2g_2)}
\ea
\ee
In other words, this Fourier transform maps from $\Z_N$ to its dual, which also happens to be $\Z_N$. We shall choose the group product on the dual $\Z_N$ to be the additive one. We define the Fourier modes in two different fashions to save us from yet more field redefinitions later. Then, the potential takes the form:
\be\label{jftb}
\ba{l}
\dsty
V_\lambda[\tilde \varphi^{(i)}] =  \frac\lambda {\sqrt{N^3}} \sum_{g}\mathfrak{Re}\left[ \tr(\tilde \varphi^{(0)}\, B^g\tilde \varphi^{(1)}\, B^g\right.\\[0.2cm]
\hspace{4.2cm}\left.\times\;\tilde \varphi^{(2)}\,B^{g}\,\tilde \varphi^{(3)}\, B^g)\right]
\ea
\ee 
where $B_{xy} = e^{\frac{2\pi i}{N}x}\delta_{x,y}$.   We now have the Boulatov model written explicitly as a matrix model.  From the redefinition of the fields \eqref{jft01}, one sees that this matrix model generates the jackets associated to the $3d$ Feynman graphs.  The diagonal shift symmetry has modified the model, in the sense that one does not have a set of $4N$ complex $N\times N$ matrices, but rather just four such matrices.  They are, however, subjected to $N$ potential terms.  The insertions $B$ are somewhat familiar from matrix models with dually weighted graphs \cite{kaz}.  These models have been used to study the statistics of branched coverings of Riemann surfaces.  While that work is not directly applicable here, it provides a new avenue to explore in group field theories.  

The peculiar dual weighting that one has here is perhaps easiest to see if one explicitly sums over $g$ in the potential \eqref{jftb}.  One loses again the ability to express the potential as a trace, but the potential now takes the form:
\be
\ba{l}
\dsty
V_\lambda[\tilde \varphi^{(i)}] =  \frac\lambda {\sqrt{N^3}} \nsum{g, g_i}\mathfrak{Re}\left[ \tilde \varphi^{(0)}_{x_1,x_2}\,\tilde\varphi^{(1)}_{x_2,x_3},\tilde \varphi^{(2)}_{x_3,x_4}\right.\\[0.2cm]
\hspace{3.7cm}\left.\times\,\tilde \varphi^{(3)}_{x_3,x_1}\delta_{x_1 + x_2 + x_3 + x_4, 0}\right]
\ea
\ee
The propagator and vertex operator look remarkably simple:
\be
\ba{rcl}
\cP_{x_i;\hat x_i} &=&\delta_{x_1,\hat x_1}\,\delta_{x_2,\hat x_2}	\\[0.2cm]
\cV_{x_i;\hat x_i} &=&\dsty\frac{\lambda}{\sqrt{N}}\delta_{x_1, \hat x_1 }\,\delta_{x_2, \hat x_2 }\,\delta_{x_3, \hat x_3 }\;\delta_{x_4, \hat x_4 }\\[0.2cm]
&&\hspace{3cm} \times\delta_{x_1 + x_2 + x_3 + x_4, 0}.
\ea
\ee
The degrees of freedom are attached to the faces of the jacket, with the constraints residing at the vertices:
\be
\cZ_{ J^{ (ij,\hat{ij}) } } = \left(\frac{ \lambda }{ \sqrt{N} }\right)^{|v_{J}|} \left[ \prod_{f^{(kl)}_J} \sum_{x_f}  \right] \prod_{v \in J} \delta\left( \sum_{f@v} x_{f}   \right)
\ee
and once again $(kl)\neq (ij)$ or $(\hat{ij})$.
 If the degrees of freedom were attached to the edges and one had a closure constraint, one could solve the constraint by a change of variables. Unsurprisingly, this is not the case here, since the amplitude captures the topology of the ambient $3d$ triangulation.  In some sense the best way to solve the constraint is to reconstruct the $3d$ manifold and perform the analysis of \cite{bfdiv}. Having said that, the power behind the reformulation is that the Boulatov model can now be expressed at the level of the action as a quantum field theory with support on the jackets. As mentioned before, there are three jackets on which one might develop a quantum field theory.  We picked one class of jacket, $J^{(02,13)}$, for the analysis above, but the other two classes are just different gauge-fixings of the Boulatov action and give the same amplitudes.

\section{Conclusions and outlook}

In this paper, we have shown that the bubbles and jackets occurring in the construction of the $1/N$-expansion are in fact generated from matrix models embedded inside the tensor model.  In fact, they constitute all the possible embedded matrix models for this potential.  

In the case of bubbles, we showed clearly that it is rather unnatural, from the tensor model point view, to hope that one can  excise a priori pseudo-manifolds from the expansion.  The case of jackets is yet more interesting.  We showed that they correspond to splitting surfaces of a handlebody decomposition of the triangulation.  When $\Delta$ is a manifold, they are indeed Heegaard surfaces.     With this property in hand, it is now possible to utilize the extensive results on Heegaard and splitting surfaces with respect to 3-manifold analysis.  To finish, we used our result to express the Boulatov model as a dually weighted matrix model on these Riemann surfaces.  

Importantly, we have now a reason to study these embedded matrix models with the array methods already available in the field of matrix models.  In coming work, we shall investigate the solvability of these models using standard matrix model techniques \cite{me2}.   Moreover, we are interested to extend the analysis from finite to Lie groups \cite{medan} since the $\SU(2)$ Boulatov group field theory has a direct relation to 3d quantum gravity.   

\section{Acknowledgements}
We would like to thank D. Oriti for fruitful discussions in the early stages of the work and B. Dittrich, D. Benedetti, V. Bonzom  and E. Livine for helpful suggestions and encouragement en-route.

\end{document}